\documentclass[12pt]{article}
\usepackage{epsfig}

\parskip=\medskipamount % space between paragraphs (TeX)
plus.3em minus.5em \abovedisplayshortskip=.5em plus.2em minus.4em
\renewcommand{\thesection}{\arabic{section}}

\catcode`@=11

\@addtoreset{equation}{section} \@addtoreset{equation}{subsection}
\def\theequation{\ifnum\value{section}=0 \arabic{equation}\ignorespaces
\else \ifnum\value{section}=-1 A.\arabic{equation}\ignorespaces
\else \ifnum\value{subsection}=0
\thesection.\arabic{equation}\ignorespaces \else
\thesection.\arabic{subsection}.\arabic{equation}\ignorespaces
                             \fi
                        \fi
                   \fi}

{\catcode`\'=\active \def'{{}^\bgroup\prim@s}}

\catcode`@=12

\newcommand{\bq}{\begin{equation}}
\newcommand{\be}{\begin{equation}}
\newcommand{\fq}{\end{equation}}
\newcommand{\ee}{\end{equation}}
\newcommand{\bqr}{\begin{eqnarray}}
\newcommand{\beqs}{\begin{eqnarray}}
\newcommand{\fqr}{\end{eqnarray}}
\newcommand{\eeqs}{\end{eqnarray}}

\newcommand{\rf}[1]{(\ref{#1})}

\def\bop#1{\setbox0=\hbox{$#1M$}\mkern1.5mu
    \vbox{\hrule height0pt depth.04\ht0
    \hbox{\vrule width.04\ht0 height.9\ht0 \kern.9\ht0
    \vrule width.04\ht0}\hrule height.04\ht0}\mkern1.5mu}
\def\Box{{\mathpalette\bop{}}}                        % box

\begin{document}
\thispagestyle{empty}

\begin{flushright}
\begin{tabular}{l}
% TEP- \\
hep-th/0505018 \\
\end{tabular}
\end{flushright}

\vskip .6in
\begin{center}

{\bf  Quantum Solution to Scalar Field Theory Models}

\vskip .6in

{\bf Gordon Chalmers}
\\[5mm]
% {\em address \\
%      address \\
% Los Angeles, CA } \\

{e-mail: gordon@quartz.shango.com}

\vskip .5in minus .2in

{\bf Abstract}

\end{center}

Amplitudes $A_n$ in $d$-dimensional scalar field theory are generated, 
to all orders in the coupling constant and at $n$-point.  The amplitudes are 
expressed as a series in the mass $m$ and coupling $\lambda$.  
The inputs are the classical scattering, and these generate, after the 
integrals are performed, the series expansion in the couplings $\lambda_i$.  
The group theory of the scalar field theory leads to an additional 
permutation 
on the $L$ loop trace structures.  Any scalar field theory, including 
those with higher dimension operators and in any dimension, are amenable.

\vfill\break 

\noindent {\it Introduction}

The quantum scalar $\phi^3$ theory has been studied for many years and 
is a textbook quantum field theory.  The interactions in this theory 
are typically examined to lowest order in perturbation theory, 
or to higher orders in the ultraviolet so as to find the critical 
exponents and scaling.  Large order studies in perturbation theory 
were performed over twenty years ago, without detailed knowledge of the 
diagrams.  The diagrams in the usual perturbtion theory are 
complicated to evaluate in general, which has  slowed progress.  

The derivative expansion has been pursued for several years 
\cite{Chalmers3}-\cite{Chalmers12}.  
This formulation has been placed in the context of many theories, including  
scalar, gauge and supersymmetric models.  The 
derivation of the quantum scattering has been simplified in \cite{Chalmers3}; 
this approach is used here to find the amplitudes of all scalar field 
theories in any dimension, to all orders in the couplings.  All coefficients 
of the following $n$-point amplitude expansion can be determined,  

\bqr 
A_n = \sum c(i,m) \lambda^i (k^2)^m \ , 
\fqr 
with $\lambda$ the coupling constant and $k^{2m}$ representing the generic 
product of the $n(n-1)/2$ momentum invariant at $n$-point.  

The tree amplitudes in $\phi^3$ theory are given in \cite{Chalmers1}.  The  
amplitudes for any scalar field theory follow from this result by pinching 
propagators. 

In general the classical amplitudes in any quantum field theory, including 
massless ones, are required to recursively construct, in this formulation, 
the solution to the amplitudes and effective action.  The recently appeared 
tree amplitudes of scalar, and gauge and gravity theory, are based on a simple 
number theoretic parameterization \cite{Chalmers1}, \cite{Chalmers2}.  These 
scalar amplitudes and their coefficients are used here in the quantum scalar 
solution.  

The classical Lagrangian that generates the amplitudes are those of massive 
scalar field theory, and includes the possible interactions.  These pertain to 
$\phi^n$, scaled with an appropriate coupling, and the derivatives $\partial^{n_1} 
\phi \ldots \partial^{n_m} \phi$.  Group theory, the inclusion of additional 
modes, and mixed interactions are also included.  The group theory adds a 
complication associated with permutations of external lines.

\vskip .2in 
\noindent {\it $n$-Point Amplitudes} 

The genus zero amplitudes are first presented.  Then the formulae 
describing the quantum amplitudes are given and used to find the full 
amplitudes.  

The tree amplitudes in $\phi^3$ at all $n$-point have recently been 
described in the literature \cite{Chalmers12}; a set of numbers $\phi_n(i)$ are required 
to specify individual diagrams.  These $n-2$ numbers label the vertices 
and range from $n$ to $1$.  In a color ordered tree, they occur at most 
$n-2$ times for the greatest number to none in the case of the lowest 
number in an incremental manner.  The set $\phi_n$ generates the 
momentum routing of the propagators and describe the diagram.

The $\phi_n$ numbers can be changed to the set of numbers $i,p$ 
which describe the poles in the diagram through the invariants 

\bqr  
t_i^{[p]}= (k_i+\ldots+k_{i+p-1})^2 \ .  
\fqr 
These invariants are defined for a fixed ordering of the external 
legs and the numbers are cyclic around the final number.  A second 
set of numbers, besides the $\sigma(i,n)$ are required when the 
mass expansion is performed.  Due to the series $1/(m^2-p^2)=m^2 \sum 
(p^2/m^2)^k$, the coefficients $\tilde\sigma(i,p)$ are numbers 
from $0$ to $\infty$ and label the exponent in the series for 
each propagator.

The numbers 

\bqr 
\sigma(i,p) \qquad \tilde\sigma(i,p) \ , 
\label{treekinematics} 
\fqr 
describe the individual diagrams in the mass expansion.

The fundamental iteration is accomplished via the sewing procedure
as described in \cite{Chalmers3}-\cite{Chalmers12}.  The integrals
are simple free-field ones in x-space, and generate an infinite
series of relations between the parameters of the coupling
expansion $\alpha^{p_{ij}}_{n,g}$. 

\bqr
\sum_q^{p_{ij}} \alpha^{p_{ij}}_{n,q} \lambda^{n-2 + q} =
 \sum_{i,j,p;l_{ij},n_{ij},m_{ij}} \alpha_{n+p,i}^{l_{ij}}
\alpha_{n+p,j}^{n_{ij}} \lambda^{2n+2p+i+j-4}
  I_{l_{ij},n_{ij}}^{p_{ij}} \ .
\label{phiiteration}
\fqr
The indices $i,j$ are exampled below.  The $\alpha$ parameters are quantum 
corrected vertex parameters, and they take into account propagator corrections.  
The coefficients  $I_{l_{ij},m_{ij}}^{p_{ij}}$ are defined by the momentum expansion
of the 'rainbow' integrals

\bqr
J_{{\tilde l}_{ij},{\tilde m}_{ij}}^{\sigma,{\tilde p}_{ij}} =
 \int \prod_{a=1}^p d^dq_a {1\over
 (q_{\rho(a)} - k_{\sigma(a)})^2 + m^2} \prod
 s_{ij}^{{\tilde l}_{ij}+{\tilde m}_{ij}} \quad\vert^{p_{ij}}
\ ,
\label{integrals}
\fqr
with ${\tilde l}_{ij}$ and ${\tilde m}_{ij}$ parameterize a subset
of the vertex lines which are contracted inside the loop.  The indices 
$\rho$ label the linear combination of the loop momenta in the internal 
lines.    

The integrals \rf{integrals} are symmetrized over the the external lines
in the formula \rf{phiiteration}; there are $n_1$ and $n_2$ external
lines on each side of the graph and $b$ parameterizes a subset of these
numbers (e.g. $n_1=1,2,3,4$, $n_2=5,6,7,8$ and $b=3,4,5,6$; the
$l_{ij}$ and $m_{ij}$ parameterize the kinematics associated with the
exernal and internal lines.  The expansion of the integral in
\rf{integrals} in the momenta generate the coefficients $p_{ij}$.
the set of numbers $\sigma(a)$ parameterize the subset of
numbers of the two vertices (forming an integral with $n$ external
lines.  The numbers $\sigma(a)$ label numbers beyond the external
lines $1,\ldots, n_1$ and $n_1+1,\ldots,n$) and are 
irrelevant because the integral is a function of their sum;
this property lends to a group theory interpretation of the final
result in terms of the coefficients
\bqr
I_{l_{ij},n_{ij}}^{\sigma,p_{ij}} \ ,
\fqr
after summing the permutations.  The numbers $i,j$ in $l_{ij}$ and
$n_{ij}$ span $1$ to $m$ (including internal lines) and those in
$p_{ij}$ span $1$ to $n$:

\bqr
l_{ij}=(l_{ij},0,\ldots,0,l_{ij})
\qquad n_{ij}=(0,\ldots 0,n_{ij},\ldots,n_{ij},
 n_{ij},\ldots,n_{ij}) \ ,
\fqr
and
\bqr
p_{ij}=(p_{ij},\ldots,p_{ij}) \ .
\fqr
This notation of $l_{ij}$, $m_{ij}$, and $p_{ij}$ is used to setup
a (pseudo-conformal) group theory interpretation of the scattering.

The details of the expansion of the integrals in \rf{integrals}
depend on the selection of the internal lines found via the
momenta of the vertices

\bqr
\lambda_{n}^{(p_{11},p_{12},\ldots,p_{nn})}
\label{momemtavertex}
\fqr
on either side of the double vertex graph.  Although the ${\tilde
l}_{ij}$, ${\tilde m}_{ij}$, and ${\tilde p}_{ij}$ depend on the
details of the contractions and sums of the lines of the
individual vertices, the actual coefficients of the iteration,
i.e. $I_{l_{ij},n_{ij}}^{p_{ij}}$, are functions only of the
vertex parameters.  The details of the expansion and the contractions
of the tensors in the integrals \rf{integrals} are parameterized by
$p_{ij}$, which label the momentum expansion of the integrals.
The coefficients $p_{ij}$ range from $0$ to $\infty$, in
accordance with the momentum expansion of the massive theory.

Although the coefficients $I_{l_{ij},n_{ij}}^{p_{ij}}$ arise from
the integral expansion, they also have a group theory description.
The dynamics of the expansion are dictated via these coefficients
for an arbitrary initial condition of the bare Lagrangian.

The iteration of the coefficients results in the simple
expression,

\bqr
 \alpha^{m_{ij}}_{n,q}  =
 \sum_{i,j,p; l_{ij}, n_{ij}} \alpha_{n+p,i}^{l_{ij}}
\alpha_{n+p,j}^{n_{ij}} I_{l_{ij},n_{ij}}^{m_{ij}} \ .
\label{coeffiteration}
\fqr 
The sums are on the number of internal lines $p$ and the powers
of the shared couplings $i$ and $j$,

\bqr
n-2+q=2(n+p)-4+i+j \qquad q=n+2p-2+i+j \ , 
\fqr
for the example of $\phi^3$.  
The numbers of momenta $l_{ij}$ and $n_{ij}$ are accorded to $s_{ij}$ 
(some of which are within the integral).
The parameters $m_{ij}$ label the external momenta, interpreted
group theoretically through the coefficient $I$.

The integrals and the iteration in \rf{coeffiteration} have to be 
performed.  The initial condition on the sum is the form of the classical 
$n$-point amplitudes, i.e $\alpha_{n,{\rm cl}}^{p_{ij}}$.  The integral 
complication is that there are invariants that: 1) contain both external 
and loop momenta, and 2) contain only loop momenta.  The sum must 
have attention to both types of invariants as the integrals are 
different with differing numbers of loop momenta.   

\vskip .2in
\noindent {\it Integrals}

The simplest integral is when all there is no tensor numerator, 

\bqr 
I_{0,0}^{m_{ij}} = \int d^dx  \prod_{a=1}^b
 \Delta(m,x) e^{ik\cdot x} \ . 
\label{notensorintegral}
\fqr  
The tensor integrals are computed via the identity, 

\bqr
 I_{l_{ij},n_{ij}}^{m_{ij}} = 
 {\cal O}^{l_{ij}}\Bigl[ s_{ij}\Bigr] {\cal O}^{n_{ij}} \Bigl[ s_{ij}\Bigr]  
 I_{0,0} ~\vert_{m_{ij}} \ , 
\label{tensorintegral}
\fqr 
which expresses all integrals via a derivative iteration on the integral 
\rf{notensorintegral}.  In this expression \rf{tensorintegral} the internal 
momenta in $s_{ij}$, with the invariants numbered by $l_{ij}$ and $n_{ij}$, 
are replaced with a differential $\partial_k$.  

The integrals are evaluated via transforming to $x$-space and using the 
Bessel form of the massive propagators.  The propagator is in $d$ dimensions, 

\bqr  
\Delta(m,x)= \lambda_d (x^2)^{-d/2+1} K_{d/2}(mx) \ ,  
\label{propagator} 
\fqr    
The parameters are left as variables to span unusual propagation, such as $\phi 
e^{-\tilde\Lambda \Box}\phi$ or $\phi \Box^2\phi$, in view of $\phi^4$ theory, 
and the 
quantization of perturbatively nonrenormalizable theories.  The integral in 
\rf{notensorintegral} evaluates to, 

\bqr 
I_{0,0}= (k^2)^{d-(L+1)(d/2-1)}\sum \beta_{a,b} \bigl({k^2\over m^2}\bigr)^a 
  \bigl({k^2\over \Lambda^2}\bigr)^b  \ .
\label{integralseries} 
\fqr 
A momentum regulator is used, and dimensional reduction is also possible.  
The coefficients in \rf{integralseries} are, 

\bqr 
\Delta(m,x)^N=(x^2)^{N\beta_1-N\beta_2/2} (m^2)^{-N\beta_2/2} \sum_{n=0}^\infty   
   {1\over n!} \sum_{a=1}^m \sum_{n_a,b_a} {}_{\vert_{\sum_{a=1}^m b_a n_a = n}}    
   {n!\over \prod_{a=1}^m (b_a n_a)!}  
\fqr 
\bqr 
x\times {\Gamma(N+1)\over\Gamma(N-m)}  \quad  c_{\{n_a,b_a\}}~ (x^2m^2)^n
\label{multipropexp}
\fqr 
with 
\bqr 
c_{\{n_a,b_a\}} = \prod_{a=1}^{b_a} \partial_u^{(b_a),n_a} \bullet 
 (x^2m^2)^{\beta_2/2} K_{\beta_2}(u)\vert_{u=m^2x^2} \ , \qquad    
  \partial_u^{(b_a),n_a} f =  \bigl(\partial_u^{(b_a)} f \bigr)^{n_a}
\label{coeffdefns} 
\fqr 
The coefficients in \rf{coeffdefns} are derived via the Bessel function series, 

\bqr 
K_{\beta_2}(u)= e^{-i\pi\beta_2/2} \bigl({-u^2\over 4}\bigr)^{\beta_2/2} 
 \sum_{m=0}^\infty 
{1\over m! \Gamma(m+1-\beta_2)} \bigl(-{u^2\over 4}\bigr)^m + 
  {\beta_2\rightarrow -\beta_2} \ , 
\fqr 
which is the contour rotated Euclidean version.  The scalar expansion 
in even powers of $x^2m^2$ is,  

\bqr 
 e^{-i\pi\beta_2/2} 2^{-\beta_2} {\Gamma(m+1) 
 \over\Gamma(m+1-\beta_2)} \bigl({1\over 4}\bigr)^m 
\fqr 
\bqr 
\hskip 1.1in 
 - e^{i\pi\beta_2/2} 2^{\beta_2} {\Gamma(m+\beta_2) 
 \over\Gamma(m+1+\beta_2)} \bigl({1\over 4}\bigr)^{m+\beta_2/2} \quad 
 m\rightarrow m-\beta_2 \ .
\fqr 
Due to the dimension $d$, there is a Taylor series for $d$ even; the 
propagator should be $\Delta_+$ and $\Delta_-$, which is expanded in 
$\Delta^N$.  Then a Taylor series expansion can be defined again for 
general $d$; else the coefficients could be used as a variant of 
dimensional reduction.  

The  parameters in \rf{coeffdefns}
are, 

\bqr  
\Delta(m,x)^N = \sum_{a=0} \delta_{N,a} m^{2(a-\beta_2/2)} 
  (x^2)^{a+N(\beta_1-\beta_2/2)} \ .  
\fqr 

The integrals are, 
\bqr 
\int d^dx e^{ik\cdot x} (x^2)^{a+N(\beta_1-\beta_2/2)} 
\fqr 
\bqr 
= (-\partial_k\cdot\partial_k)^{-a} {\Gamma(-N(\beta_1-\beta_2/2)+1-a)\over 
  \Gamma(-N(\beta_1-\beta_2/2)+1)} 
 \int d^dx e^{ik\cdot x} (x^2)^{N(\beta_1-\beta_2/2)}
\fqr 
\bqr 
= \Bigl({\Gamma\over\Gamma}\Bigr) 
 (\partial_k\cdot\partial_k)^{-a} \rho(\beta_1,\beta_2) 
  (k^2)^{-d/2-N(\beta_1-\beta_2/2)}
\fqr 
\bqr  
\hskip -.3in
= \rho(\beta_1,\beta_2) 
{\Gamma\bigl(-N(\beta_1-\beta_2/2)+1-a\bigr) \over 
                      \Gamma\bigl(-N(\beta_1-\beta_2/2)+1\bigr)} 
\fqr 
\bqr 
\times 
   {\Gamma\bigl(-d/2-N(\beta_1-\beta_2/2)+1\bigr)\over 
     \Gamma\bigl(-d/2-N(\beta_1-\beta_2/2)+1-a\bigr)}  
  (k^2)^{-d/2-N(\beta_1-\beta_2/2)-a} \ , 
\label{beta}
\fqr 
with 

\bqr 
\rho(\beta_1,\beta_2) =(-1)^a \int d^dx e^{ik\cdot x} (x^2)^{N\beta_1-N\beta_2/2} 
\fqr 
The coefficients are defined dimensional reduction and momentum (string-inspired) 
cutoff and to find the theory in multiple dimensions.  

The tensor integrals have been examined in \cite{Chalmers8}.  Momentum of plane 
waves in x-space, within the effective action have the form of a derivative, as in 
quantum mechanics: $k_i \leftrightarrow i\partial_i$ via the Fourier transform.  
The integrals with the internal derivatives could be evaluated directly.

Those in \rf{integrals} have the form \rf{tensorintegral}, with the 
internal derivatives (or momenta) extracted from the internal lines; these momenta 
have an action on the integral in \rf{notensorintegral} as, 

\bqr 
\partial^\mu \Delta(m,x)^N = [(2-d)+m^2\partial_m^2] \times 
 \bigl( {x^\mu\over x^2}\bigr) \Delta(m,x)^N \ .  
\label{massdifferential}
\fqr 
The number $n$ counts the $\partial_x$'s.  
The $x$ factors are removable via, 

\bqr 
(\partial_{k_\mu})^{-1} = -i{x_\mu\over x^2} \ , 
\label{xextraction}
\fqr 
so that the general tensor integral requires only the scalar evaluation, followed by 
tensor derivatives as in \rf{tensorintegral}.  

The derivatives in \rf{massdifferential} have the effect of changing $\delta$ to 

\bqr 
\delta_{N,a} \rightarrow 
 \delta_{N,a} ~\rho \sum_{l=0}^n {n!\over l!(n-l)!} (2-d)^{n-l} 
 {\Gamma(a-N\beta_2/2+1)\over\Gamma(a-N\beta_2/2+1-l)} .
\label{tensorsimplified}
\fqr 
The $\delta$ changes, due to the $m^2\partial_{m^2}$ differential operators, if 
$l\geq a-N\beta_2/2$.  If $l< a+N\beta_2/2$ then $\delta=0$.  

The action of the inverse derivatives in \rf{xextraction} on 
\rf{tensorsimplified} further modifies the $\delta$ to, 

\bqr 
\delta_{N,a} (k^2)^{-d/2-N(\beta_1-\beta_2/2)-a} \rightarrow \delta_{N,a}  
  {\Gamma\bigl(d/2+N(\beta_1-\beta_2/2\bigr)+a+1-n)\over 
   \Gamma\bigl(d/2+N(\beta_1-\beta/2)+a+1\bigr)}  
\fqr 
\bqr 
\times \prod_{j=1}^n \partial_{k_{\mu_j}} (k^2)^{-d/2-N(\beta_1-\beta_2/2)-a-n} \ . 
\fqr 
The latter tensor is, 

\bqr 
\prod_j k_{\mu_j} ~ 2^n {\Gamma(-d/2-N(\beta_1-\beta_2/2)-a-n+1) 
                  \over\Gamma(-d/2-N(\beta_1-\beta_2/2)-a-2n+1)} 
                 (k^2)^{-d/2-N(\beta_1-\beta_2/2)-a-2n} 
\label{tensorone} 
\fqr 
\bqr 
\sum_{\rm perms} \eta_{\mu_1\mu_2} \prod_j^{n-2} k_{\mu_j} ~ 2^{n-1} 
   {\Gamma(-d/2-N(\beta_1-\beta_2/2)-a-n+1) 
     \over\Gamma(-d/2-N(\beta_1-\beta_2)-a-2n+2)} 
\fqr 
\bqr 
\times   (k^2)^{-d/2-N(\beta_1-\beta_2/2)-a-2n+1} 
\label{tensortwo}
\fqr 
and on, via incrementing the factorial and multiplying the number of metrics when 
the number of derivatives is even.  The general form is, 

\bqr 
\sum_{\sigma_w,\tilde\sigma_w} \prod_{i=1}^w 
  \eta_{\mu_{\sigma(i)}\mu_{\tilde\sigma(i)}} 
    \prod_{i=1}^{n-w} k_{\mu_{\rho(i)}} ~  
  2^{n-w} {\Gamma(-d/2-N(\beta_1-\beta_2/2)-a-n+1)\over 
       \Gamma(-d/2-N(\beta_1-\beta_2/2)-a-2n+1+w/2)} \ . 
\label{generaltensor} 
\fqr 
with a factor $(k^2)^{-d/2-N(\beta_1-\beta_2/2)-a-2n+w/2}$
The $\sigma$ and $\tilde\sigma$ are vectors with $w$ components, and there is 
a summation over all combinations.  The $\rho$ set is the complement of 
these two vectors in the space of the $n$ components.

The net result for the tensor integrals is, 

\bqr 
\int d^dx e^{ik\cdot x} \prod^n \partial_{\mu_j} 
 \Delta(m,x)^N = T_{\mu_j}^n \sum_{a=1}^\infty \delta(N,a) 
  (m^2)^{a-N\beta_2/2} (k^2)^{-d/2-N(\beta_1-\beta_2/2)-a}  
\fqr 
 
\bqr 
= \gamma i^n \sum_{p=0}^\infty   
   {1\over p!} \sum_{p_a,b_a} {}_{\vert_{\sum b_a p_a = p}}    
   {p!\over \prod_{p=1}^m (b_a p_a)!}  
 {\Gamma(N+1)\over\Gamma(N-m)} ~c_{\{b_a,p_a\}}
\fqr 
\bqr 
\rho(\beta_1,\beta_2) 
{\Gamma\bigl(-N(\beta_1-\beta_2/2)+1-p\bigr) \over 
                      \Gamma\bigl(-N(\beta_1-\beta_2/2)+1\bigr)}
{\Gamma\bigl(-d/2-N(\beta_1-\beta_2/2)+1\bigr)\over 
                      \Gamma\bigl(-d/2-N(\beta_1-\beta_2/2)+1-p\bigr)}  
\fqr 
\bqr
\rho \sum_{l=0}^n {n!\over l!(n-l)!} (2-d)^{n-l} 
 {\Gamma(p-N\beta_2/2+1)\over\Gamma(p-N\beta_2/2+1-l)} (m^2)^{p-N\beta_2/2} 
 (k^2)^{-d/2-N(\beta_1-\beta_2/2)-p}  
\fqr 

\bqr 
\times {\Gamma\bigl(d/2+N(\beta_1-\beta_2/2\bigr)+p+1-n)\over 
   \Gamma\bigl(d/2+N(\beta_1-\beta/2)+p+1\bigr)} 
\fqr 

\bqr 
\sum_{\sigma_w,\tilde\sigma_w} \prod_{i=1}^w 
  \eta_{\mu_{\sigma(i)}\mu_{\tilde\sigma(i)}} 
    \prod_{i=1}^{n-w} k_{\mu_{\rho(i)}} ~  
  2^{n-w} {\Gamma(-d/2-N(\beta_1-\beta_2/2)-p-n+1)\over 
       \Gamma(-d/2-N(\beta_1-\beta_2/2)-p-2n+1+w/2)}
\label{wintegrals}
\fqr 
The sums on $l$ and $a$ should be performed, in order to have a simplified 
expression at fixed tensor structure.  The momentum $k$ is the sum of the 
momenta on the exterior of the integral, i.e. $k=\sum_{j=1}^q k_j$.  The 
number $n$ refers to the number of derivatives $\partial_\mu$ on the 
internal lines of the integral.  $m$ referes to the maximum number 
$a$ in $b_a$ and $p_a$, except where the mass term is obvious.    

\begin{figure}
\begin{center}
\epsfxsize=8cm
\epsfysize=8cm
\epsfbox{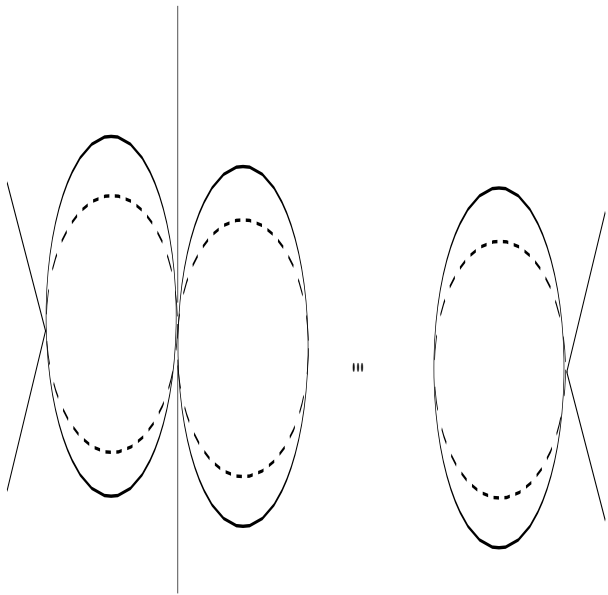}
\end{center}
\caption{The diagram showing the iteration to the recursive formula.  This figure 
shows how tree diagrams are used to construct loop amplitudes, when both are momentum 
expanded.  The recursion is not required from this point of view.  The internal and 
external lines within the loop are to be an indefinite number.  }  
\end{figure}

\vskip .2in 
\noindent {\it Solution to Coefficients}

The iterative formula in \rf{coeffiteration} can be expanded into a product form.  
The substitution of the prior $\alpha$ terms into the expression will continue 
until the the $\alpha_{q;i}^{q_{ij}}$ represents the classical scattering.  In 
$\phi^3$ theory this occurs at $q-2=i$ coupling order ($q-2$ vertices); more 
general scalar theories have more than one coupling constant and the count is 
more complex.  

The expansion of the iterative formula is represented in Figure 1; the sum of 
nodes from $2$ to $n_{\rm max}$ is required.  The number of an individual 
propagator can be any integer, in conjunction with the expansion of a tree 
diagram.  The external legs are permuted at the nodes appropriate to the color 
structure.  

The tree level initial conditions are required to solve the recursion; a scalar 
field theory possessing higher derivative terms can model any initial condition.  
The $\phi^3$ and $\phi^4$ initial conditions are described in \cite{Chalmers1}, with a 
bootstrap condition $q-2=i$ and $q-2=2i$; $q$ is the external leg number and 
$i$ counts either the 3- or 4-point vertices (coupling constants).

The recursion solution is, 

\bqr 
\alpha_{n,q}^{n_{ij}} = \sum_{a_{\rm nodes}=1}^{N_{\rm nodes}} \sum_{ n^{(c)},n^{(c)}_{ij} }
\quad \prod_{a=1}^{a_{\rm nodes}}  \alpha_{n^{(a)},q^{(a)}}^{n^{(a)}_{ij}} 
\prod_{b=1}^{a_{\rm nodes}-1} I_{n^{(b)}_{ij},n^{(b+1)}_{ij}}^{n^{(b)}+n^{(b+1)}} \ , 
\label{productform}
\fqr 
with $b_{\rm nodes}=a_{\rm nodes}-1$.  The number of nodes is to be summed; the maximum 
is set by the initial conditions.  External lines may exit from any of the nodes.  The 
numbers of propagators have to be summed at each of the nodes, when $a_{\rm nodes}\geq 
2$.  The $\alpha_{n^{(a)},q^{(a)}}^{n^{(a)}_{ij}}$ parameters are classical (loop zero); 
different boundary counditions could iterate from non-classical data without altering 
the form of \rf{productform}.  When there is more than one coupling constant, $q=\sum q^{(a)}$.  The classical coefficients in $\phi^3$ theory are  

\bqr 
\alpha_{n^{(a)},q^{(a)}}^{n^{(a)}_{ij}} =  (m^2)^{n^{(a)}-3} \lambda^{n^{(a)}-2} 
  \prod (m^2)^{-n_{ij}^{(a)}} \ , 
\fqr 
with the latter factor representing the mass expansion of the propagators.  

The parameters $n^{(a)}_{ij}$ project the form in \rf{integrals} at a fixed 
tensor.  At each node there are $n_{b-1}$ lines to the left and $n_{b}$ lines 
to the right.  The tensor structure is denoted by $n^{(b)}_{ij}$, and all 
of the kinematics of the $s_{ij}$ in the integrals add to form the tensor 
of $\alpha_{n,q}^{n_{ij}}$.  

In order to find the product of the integrals at fixed parameters, in 
\rf{integrals}, the kinematics at the vertex are expanded as, 

\bqr  
s_{ij}=(k_i+i\partial_x)^2 = 2i k_i\cdot\partial_x - \partial_x^2 \ ,
\label{eilegs}
\fqr 
and 
\bqr 
s_{ij}=-4\partial_x^2 \ . 
\label{iilegs} 
\fqr 
The first example is the situation when $i$ is an external leg and the other 
line is an internal leg; the latter has both internal legs.  The 
derivatives are those in \rf{massdifferential}.  Starting at the left 
node, the number of momenta which are internal are counted so as to define 
the tensor in \rf{integrals}.  The numbers $\phi_n$ are used for this 
count \cite{Chalmers1},\cite{Chalmers2}.  These numbers are a (symmetric) 
set theoretic foundation to build any $\phi^3$ tree diagram.  Given the 
set $\phi_n$, a function has has to be made that counts the number of 
$s_{ij}$ belonging to the internal-external (ie) class.  Else the explicit 
tree diagram labeled by the numbers of $n_{ij}$ has to be used, without 
the simple set theoretic definition. 

At the node, there is a tensor from the expansion of the invariants in 
\rf{eilegs} and \rf{iilegs}.  The number of spatial derivatives ranges 
from $\Gamma_1(\phi_n,n_{ij},\upsilon)$ to $\Gamma_2(\phi_n,n_{ij},\upsilon)$. 
These numbers depend on the external leg set $\upsilon$ and the set of 
$s_{ij}$ as defined by $\phi_n$ and $n_{ij}$.  The difference between 
the two numbers is due to the counts $s_{ii}$ and $s_{ie}+s_{ii}$, 
i.e. the number of invariants with shared legs.  The sets $\phi_n$ are 
not required if the input $n_{ij}$ is given independently.  To each 
of these counts is a tensor $W_{\nu_j;n}$.

Each of the integrals has $p_{(b)}$ internal lines.  The tensors $W_{\nu;n}$ 
contract with the momentum of the external line as the integral is a function 
only of $k$; the explicit form is in \rf{wintegrals}.  If there are 
external legs attached to the node, as illustrated on node 2 in Figure 2, 
then $k= \sum_{\rm node b} k_j=\sum_{\rm node i} p_i$; these momenta 
contract with the node tensors $W_{\nu;n}$.

The expansion of the invariants 

\bqr 
(-4\partial_x^2)^{s_{ii}} 
\prod_{i,j\in ie} (2ik_i\cdot\partial_x-\partial_x^2)^{n_{ij}}  
 = T^{(b)}_{\mu_i,w,k_j} \prod \partial^{\mu_i}  \ , 
\label{innerproducts} 
\fqr 
defines the tensors $T_{\mu,w,k_j}$ for the variable $w$ and node $b$. 
The node momentum $p=\sum_a^b p_a$ is a sum of the previous on-shell 
momenta $k_{\sigma(m_i-m_b)}$ to $k_{\sigma(m_f+{\tilde m}_b)}$.  As 
a result these invariants are expressed in terms of the two-particle 
invariants through $p^2=\sum_{i<j} s_{{ij}}$; these $s_{ij}$ variables 
are used to define the tree and the adjacent loop integrations.

The tensor in \rf{innerproducts} is 

\bqr 
T^{(b)}_{\mu_i,w,k_j}= C(q_1) (-4)^{q_2} (2i)^{q_1} 
 \prod_{i=1}^{q_1} k_{\alpha_\sigma(i)}
 \prod_{i=1}^{q_2} \eta_{\nu_i\nu_{(i+1/2)}} 
 \prod_{i=1}^{q_1} \eta_{\alpha_{\sigma(i)}\nu_i} 
\fqr 
with the prefactor $C(q_1)$ defined from \rf{innerproducts} 

\bqr 
C(q_1) = \prod_{ij\in ie} 
  {n_{ij}!\over (n_{ij}-{\tilde n}_{ij})! {\tilde n}_{ij}!} 
   (-1)^{(n_{ij}-{\tilde n}_{ij})}
\qquad q_1=\sum {\tilde n}_{ij} \ . 
\fqr 
The remaining 
derivatives contract with the external momenta set $\sigma(i)$.  The loop 
tensor from \rf{integrals} is 

\bqr 
\sum_{\sigma_w,\tilde\sigma_w} \prod_{i=1}^w 
  \eta_{\mu_{\sigma(i)}\mu_{\tilde\sigma(i)}} 
    \prod_{i=1}^{n-w} k_{\mu_{\rho(i)}} \ . 
\label{looptensor}
\fqr 
The contraction of the two tensors results in 

\bqr 
T(k_i)= \sum_{\beta,\tilde\beta,\alpha} B_{d_1} ~
   (P_b^2)^c \prod s_{\beta(i)\tilde\beta(i)} \prod k\cdot k_{\alpha(i)} \ ,
\label{tensorcontract}
\fqr  
with $\beta(i)$ and ${\tilde\beta}(i)$ denoting labels in the set of 
indices $\sigma(m_i-m_b)$ to $\sigma(m_f+{\tilde m}_b)$.  The $(P_b^2)^c$ 
is expanded to the terms, 

\bqr 
(\sum s_{ij})^c = \sum_{\rho,\tilde\rho}  
  \prod_{k}^c s_{\rho(k){\tilde\rho}(k)} \ ,   
\label{kexpansion}
\fqr 
with $\sigma$ and $\tilde\sigma$ specifying the permutations in the product 
from the power $c$; $k$ range from $1$ to $c$.  The numbers $i$ and $j$ are 
pairs of numbers between $\sigma(m_i-m_b)$ and $\sigma(m_f+{\tilde m}_b)$, 
which represent the indices of the external momenta at node b.  The permutation 
sets are all combinations of the pairs of numbers $i,j$ including repeating pairs, 
which is the same as all sets of numbers $i$ and $j$ including repeating 
ones.  

The last term in \rf{tensorcontract} in terms of two-particle invariants 
is, 

\bqr 
\sum_{\tilde\alpha}
\prod_{i,j} s_{k_{{\tilde\alpha}(j)} k_{\alpha(i)}} \ . 
\fqr 
The $\tilde\alpha$ is summed over all combinations of the momentum 
labels in the set of $P_b$, the momentum flowing into the loop at 
node $b$.  The expansion is then all pairs of numbers ${\tilde\alpha} 
=\sigma(m_i-m_b), \ldots, \sigma(m_f+{\tilde m}_b)$ and $\alpha$ 
with the first set repeating in all possible ways.  

The net result for the tensor at level $w$ is a collection of 
$s_{ij}^{n_{ij}}$.  These two-particle invariants are all external 
lines to the loop system at node $b+1$.  The number of these 
invariants is denoted $m_{ij}^{n_{b-1}}$, which is a function 
of the preceeding node $b-1$ and the number of propagators.    

The integral factors in the formula \rf{productform} multiply the 
tensor products in \rf{innerproducts}.  These functions are the 
product 

\bqr 
\alpha_{n,q}^{n_{ij}} = 
\sum_{\{n_b\}} \prod_{b=1}^{b_{\rm max}}I_{n_{(b)},w_{b}} \quad T(k_i) \ ,   
\fqr 
\bqr 
T_{(k_i)} = \prod_{b=1}^{\rm nodes} \prod_{i,j} s_{ij}^{m_{ij},n_{b-1}} 
  \prod_{i,j\in ee,b_{\rm max}} s_{ij}^{m_{ij}} 
\fqr 
The integral product is found from multiplying the scalar integrals 
in \rf{integrals}; these depend on the number of propagators and the 
index $w$.  The summation over the propagators is independent at 
each integral, which does change the initial condition 
$\alpha_{n,q}^{n_{ij}}$.  The $n_{b-1}$ is the propagator number at 
node $b$, counting lines to the left; at $b=1$, $m_{ij}$ counts the 
external-external invariants.

Having found the scalar product and the tensor product, the bounds 
on the sum require to be defined.  The coupling order $q$ can 
partition into the nodal orders via $q=\sum_b^{\rm nodes} q_b$; 
the minimum and maximum for $\phi^3$ theory is $q_b=2$ and $q_b=q-2$.
The number of partitions is 

\bqr 
\sum_{b=2}^{b_{\rm max}}
\sum_{\{q_m\};b_{\rm nodes}} {q!\over q_1!q_2!\ldots q_b!} 
 = \sum_{b=2}^{\rm max}  {q!\over b!} b^b  - {\rm boundary~terms}
\fqr 
with the boundary conditions appropriate to the theory; the latter remove 
th $q_b=1,0$ and $q_b=q-1,q$.

The permutation sum on the external lines has to be performed.  The $n$ 
external lines are to be placed in all possible ways located at the $b$ 
nodes. The trace structures of a non-abelian theory require the permutation 
subsets of the external lines; the color flow appears simpler due to the 
topology of the rainbow graphs.  

The iterative formulae has to include the propagator corrections.  It 
appears that these corrections were not included in the product form 
of the amplitudes.  However, the quantum vertices in \rf{coeffiteration} 
take into account these quantum corrections, and so should the latter 
form.  Tree diagrams in the mass expansion allow the generation of 
propagator corrections

(If there is a formal reason to examine the amplitudes without the propagator 
corrections, or $m$-point corrections, then it is possible to extract them.
This is done by modifying the external lines of the tree amplitudes with the 
mass expansion of the full quantum two-point function, i.e. $\sum T_p 
m^{2p} \Box^p \Delta$.  Each of the $\Box$s on the internal lines is 
included in the iteration and the $T_p$ coefficient modifies the vertex.
The summation on the derivatives in the classical vertex takes into 
account the propagator correction by altering the limits on the classical 
vertex and multiplying the $T_p$; there is a $p_j$ on each internal line 
that modifies the count of $n$ by $n\rightarrow n+\sum 2p_j$.  The vertex 
gets a factor $\prod T_{p_i}$.  To eliminate the propagator correction, at 
each vertex divide by the $T_{p_i}$ numbers and alter the sum by $n\rightarrow 
n-\sum 2p_j$.  The $m$-point corrections are eliminated in a similar fashion.)

\vskip .2in 
\noindent {\it Concluding remarks} 

The quantum theory to any scalar field theory is generated in a direct 
manner; $n$-point scattering amplitudes are composed through a product 
of tree amplitudes of varying coupling orders and with varying numbers 
of legs.  The conservation of both is used to find the coefficient of 
the scattering at any loop order $L$ and $n$-point.  Formulae in this 
paper demonstrate a simple sum of products which arise from a breaking of 
these orders into partitions; an example is $\phi^3$ theory in which 
the coupling order is $m-2$ with $m$ being the number of legs.  All scalar 
field theories, including those models with any number of higher 
dimensional operators, are 
quantized with the initial condition of the classical scattering.  
The solutions  could lead to better formulations of the quantum 
scalar models and the possible geometries, or potential conformal models, 
generating them; this includes the large coupling regime.

The requirements for any of these massive theories to be solved at any order 
are the tree amplitudes, which are expanded in low energy (e.g. see 
\cite{Chalmers1}).  The formulae involving the $\Gamma$ summations and the 
tensors can be simplified, and this will will lead to a more compact representation, 
less than this the half page of algebra given in this work.  

The unitarity has been made indiscreet due to the large mass or low momentum 
expansion.  This can be found by resumming the momentum modes at a specific 
order.  This has been examined in \cite{Chalmers3},\cite{Chalmers7},\cite{Chalmers8},\cite{Chalmers10},\cite{Chalmers12} in the this expansion.

This work shows that the solution to a quantum field theory can be obtained, 
in the case of an arbitrary scalar field theory in any $d$-dimension.  The 
same is true for gauge and gravity theories (for which the tree amplitudes 
have been found in a number context in \cite{Chalmers2}, and also in the 
standard model.  The formulae are similar but with more complicated 
tensor algebra. 

The complete scalar classical scattering in generic non-linear $N=2$ 
sigma models also follows number theoretically from the classical 
$\phi^3$ theory, and this includes all toric Calabi-Yau quotients  
\cite{ChalmersInPrep}; the complete quantum solution 
of these scalar models can be found.   

{\it Note added:} There are further simplifications of the tensor 
algebra.  The index and integral forms are suitable for a computer
implementation of the scattering derivation.  Revision to text is 
notation and a gamma function in equations (38)-(43).     

The form of amplitude is, 

\bqr 
A_n = g^L \prod s_{ij}^{n_{ij}} \sum \prod B(n_i,m_i,w_i) \ , 
\fqr 
with $n_i$ and $m_i$ parameterizing the number of internal lines and 
derivatives acting on them, and $w_i$ the number of metric tensors.  
The notation follows equation \rf{wintegrals}.  The sum on these numbers 
depends on the number of loops $L$ and the kinematics $n_{ij}$.  The 
functions $B(n,m,w)$ are the 'building blocks', the sums given in 
\rf{wintegrals}.  

\vfill\break 
 
\end{document}